\documentclass[a4paper, 11pt]{amsart}
\usepackage{graphicx}
\usepackage{placeins}
\usepackage{float}
\usepackage{subfigure}
\usepackage{verbatim}
\usepackage{amsmath, amssymb, amsthm, url}%, wasysym}
\usepackage[english]{babel}
\usepackage[T1]{fontenc}
\newcommand{\R}{\mathbb R}

\def\be#1\ee{\begin{equation}#1\end{equation}}
\newcommand{\fer}[1]{(\ref{#1})}
\newtheorem{theorem}{Theorem}[section]

\newtheorem{remark}[theorem]{Remark}

\newcommand{\bq}{\begin{equation}}
\newcommand{\eq}{\end{equation}}
\newenvironment{equations}{\equation\aligned}{\endaligned\endequation}
\def\bqa{\begin{eqnarray}}
\def\eqa{\end{eqnarray}}

\def\e{\epsilon}

\def\t{\theta}

\newcommand{\bd}{\begin{displaymath}}
\newcommand{\ed}{\end{displaymath}}
\newcommand{\ba}{\begin{eqnarray}}
\newcommand{\ea}{\end{eqnarray}}

\def\R{\mathbb{R}}

\theoremstyle{plain}

\title{Call center service times are lognormal. A Fokker--Planck description}

\author{Stefano Gualandi and Giuseppe Toscani}

\thanks{Department of Mathematics  of the University of Pavia, Italy.  e.mail: stefano.gualandi@unipv.it, giuseppe.toscani@unipv.it. 
}

\date{\today}

\begin{document}
\maketitle

\begin{center}\small
\parbox{0.85\textwidth}{

\textbf{Abstract.} Call centers are service networks in which agents provide telephone-based services. An important part of call center operations is represented by service durations. In recent statistical analysis of real data, it has been noticed that the distribution of service times reveals a remarkable fit to the lognormal distribution.
In this paper we discuss a possible source of this behavior by resorting to classical methods of statistical mechanics of multi-agent systems. The microscopic service time variation leading to a linear kinetic equation with lognormal equilibrium density is built up introducing as main criterion for decision  a suitable value function in the spirit of the prospect theory of Kahneman and Twersky.
\medskip

}
\end{center}

\section{Introduction}

Call centers constitute an increasingly important part of business world, employing millions of
agents across the globe and serving as a primary customer-facing channel for firms in many
different industries. For this reason, many issues associated with human resources management, sales, and marketing have also become increasingly
relevant to call center operations and associated academic research. Mathematical models are built up by taking into account statistics concerning system
primitives, such as the number of agents working, the rate at
which calls arrive, the time required for a customer to be served,
and the length of time customers are willing to wait on hold
before they hang up the phone and abandon the queue. Outputs
are performance measures, such as the distribution of time
that customers wait \emph{on hold} and the fraction of customers that
abandon the queue before being served \cite{AAM,Brown}.

In this note, we will be mainly concerned with the mathematical modeling of the duration of service times, namely with the time required for customers to be served. A deep inside into service times in a call center with hundreds of operators was the object of a recent statistical research  reported by one of the present authors to an  Italian telephonic company. The analysis of the huge amount of data provided by the company, covering a one-month period, made evident the almost perfect fitting of the statistical distribution of service times to a lognormal one. Among others, the same phenomenon was noticed before in \cite{Brown}, even if the conclusion there was that the true distribution is very close to lognormal, but is not exactly lognormal. It is remarkable that lognormal distributions appear in many biological, physical and socioeconomic phenomena, and are well studied from many years now. The interested reader can learn about by reading the books \cite{Ait,Cro} (cf. also the recent exhaustive review paper \cite{Lim}).

The discussions about the reasons behind the forming of a lognormal (steady) distribution in agents service times lead to (at least for us interesting) question to investigate if this behavior could be explained by means of the well-consolidated methods of statistical mechanics and the consequent kinetic modeling. Indeed, in recent years, there has been an increasing interest in developing kinetic models able to describe formation of macroscopic universal behaviors in a multi-agent society, by resorting to methods typical of statistical mechanics \cite{NPT,PT13}.  Recent contributes to this powerful mathematical approach, in which behavioral aspects of agents constitute a relevant aspect \cite{Per}, have been obtained by Bellomo and his coworkers \cite{ABG,BHT,BKS,BCKS,BDG}. 

 Among others, an important issue, which reveals unexpected similarities with the present problem,  was to understand the statistical behavior of agents acting in a simple financial market, by taking into account specific behavioral aspects. A kinetic approach  has been proposed in \cite{MD}. This research studies a relatively simple kinetic model for a financial market characterized by a single stock or good and an interplay between two different trader
populations, chartists and fundamentalists, which determine the price dynamics of the stock. The model has been inspired by the microscopic Lux--Marchesi model \cite{LLS,LMa,LMb}. The financial rules depends here from the opinion of traders through a kinetic model of opinion formation recently introduced by one of the present authors in \cite{To1}. Also, some psychological and behavioral components of the agents, like the way they interact with each other and perceive risks, which may produce non-rational behaviours, were taken into account. This has been done by resorting to a suitable \emph{value function} in agreement with the prospect theory by Kahneman and Twersky \cite{KT,KT1}. 

A careful reading of the mathematical approach in \cite{MD} makes evident the basic role of the human behavior, which is reproduced by the nonlinear value function, in the microscopic mechanism leading to the underlying kinetic equation, able to reproduce the macroscopic evolution of the market. Also, \emph{mutatis mutandis}, it suggests how the presence of a suitable value function can justify at a microscopic level the mechanism of formation of the service time distribution. 

Having in mind the strategy of \cite{MD}, in  Section \ref{model}  we will introduce a linear kinetic equation of Boltzmann type \cite{PT13}, which describes the distribution of service times on the basis of a simple microscopic interaction mechanism, which aims in reproducing the decision of agents about forthcoming services in presence of risks, which are here represented by the unpleasant possibility that the actual service  has been done by employing more than the expected mean time fixed by the firm to close the operation. The statistical nature of the service time is clear by considering that the number of operation performed  by the agents is of the order of half a million in a month. 

In Section \ref{quasi} we shall show that the kinetic model gives in a suitable asymptotic
limit (hereafter called quasi-invariant service time limit) a partial differential equation
of Fokker-Planck type for the distribution of time service among agents. 
The equilibrium state of the Fokker-Planck equation can be computed explicitly
and is given by a lognormal density with mean related to the expected mean time for service fixed by the firm. Similar asymptotic analysis were performed on a kinetic model for a simple market economy with a constant growth mechanism \cite{CPP,DMTb}, showing formation of steady states with Pareto tails, for kinetic equations for price formation \cite{TBD}, and in the context of opinion formation \cite{To1}. A general view about the asymptotic passage from  kinetic equations based on general interactions and Fokker--Planck type equations can be found in \cite{FPTT}.

Last, in Section \ref{numerica} we discuss the validity of the Fokker--Planck approach to the formation of a lognormal profile by resorting to the numerical fitting of a set of real data, which contain the detailed description of a one month activity of a call center. 

%%%%%%%%%%%%%%%%%%%%%%%%%%%%%%%%%%%%%%%%%%%%%%%%%%%%%%%%%%%%%%%%%
\section{A kinetic model for service time forming}\label{model}

On the basis of statistical mechanics, to construct a model able to study the evolution of the time service distribution, the fundamental assumption is that agents are indistinguishable \cite{PT13}. This means that an agent's state at any instant of time $t\ge 0$ is completely characterized by his service time duration $w \ge0$.
The unknown is the density (or distribution function) $f = f(w, t)$, where $w\in \R_+$ and the time $t\ge 0$. Its time evolution is described, as shown later on, by a kinetic equation
of Boltzmann type.
The precise meaning of the density $f$ is the following. Given the system of agents to study, and given an interval or a more complex sub-domain $D  \subseteq \R_+$, the integral
\[
\int_D f(w, t)\, dw
\]
represents the number of individuals with  characterized by a service time  $w \in D$ at time $t > 0$. It is assumed that the density function is normalized to 1, that is
\[
\int_{\R_+} f(w, t)\, dw = 1.
\]
As always happens when dealing with a kinetic problem in which the variable belongs to 
the domain $\R_+$, one is faced with supplementary mathematical difficulties in the correct
definition of \emph{interactions}. In fact, it is essential to consider only interactions that do not produce negative service times. 
In order to build a realistic model, this limitation has to be coupled with a reasonable
physical interpretation of the process of service time forming. In other words, the impossibility of crossing the boundary at $w=0$ has to be a by-product of good modeling of interactions.

Let us now proceed to describe the characteristic \emph{interaction} of the model, namely the change in future time serving of any agent who employed a certain time to  conclude its work in a single operation. The starting assumption is that agents  have precise instructions from the service manager to conclude the service in a certain ideal time $\bar w$, in any case remaining below a certain limit time for the service, denoted by $\bar w_L$, with $\bar w_L > \bar w$. It is reasonable to assume that most agents will take the value $\bar w_L$ (instead of $\bar w$) as the significant time to respect. Indeed, agents will be really afraid of having consequences about their delays in serving only when the service time is above the value $\bar w_L$.  Consequently, if an agent concluded a service in a time $w > \bar w_L$, he will accelerate  to conclude its forthcoming service in a shorter time. Likewise, in the case in which the agent was so quick (or lucky) to conclude a service in a time $w < \bar w_L$, he will work  leisurely to conclude its forthcoming service in a better way, by using a longer time. It is clear that the two situations are completely different, since in the former an agent will be worried about its working situation, while in the latter the agent will be relaxed. This is the classical situation excellently described by Kahneman and Twersky in \cite{KT}, a pioneering paper devoted to describe decision under risk. Inspired by this idea, we will describe an agent's modification of time service as
 \be\label{coll}
 w_* = w - \Phi(w/\bar w_L) w + \eta w.
 \ee
 The function $\Phi$ plays the role of the \emph{value function} in the prospect theory of Kahneman and Twersky \cite{KT}. 
However, at difference with a classical value function, which is positive and concave above the reference value $1$ ($w > \bar w_L$), while negative and convex below ($w < \bar w_L$),  we hypothesize that in the present situation the value function for changes of time services is a bounded and concave function,  positive above the reference value $1$ ($w > \bar w_L$), while negative  below ($w < \bar w_L$).   We will assume as value function with these characteristics
 \be\label{vf}
 \Phi(s) = \gamma \frac{s^\delta -1}{s^\delta +1} , \quad  s \ge 0.
 \ee
In \fer{vf} $0<\gamma < 1$ and $0 < \delta < 1$ are suitable constants characterizing the agents behavior. In particular, the value $\gamma$ will denote the maximal amount of change in service time that agents will be able to perform in a single operation. Note indeed that the value function $\Phi(s)$ is such that 
 \be\label{bounds}
  -\gamma \le \Phi(s) \le +\gamma.
 \ee
 
The presence of the minus sign in front of the value function $\Phi$ is due to the obvious fact that an agent will be induced to increase its working time when $w < \bar w_L$, and to decrease it if $w >\bar w_L$. Note moreover that the function $\Phi(s)$ is such that, given  $0 < s < 1$
 \[
 - \Phi\left(1-s \right) > \Phi\left(1+s \right).
 \]
Therefore, given  two agents starting at the same distance from the prescribed limit service time $\bar w_L$ from below and above, it is easier for the agent starting below to move closer to the optimal  service time, than for the agent starting above. Indeed, the perception  an agent will have of his work will be completely different depending of the sign of the value function. 

Last, to take into account a certain amount of unpredictability in any future realization of the service, it is reasonable to assume that any forthcoming time $w_*$ can have random variations (expressed by $\eta w$) which in the mean are negligible, and in any case are not so significant to produce a sensible variation of the service time $w$. Also, to be consistent with the necessary positivity of the service time $w_*$, it is assumed that the random variable $\eta$ takes values in the interval $(-1 +\gamma, +\infty)$, while $\langle \eta\rangle = 0$. Here and after, $\langle \cdot \rangle$ denotes mathematical expectation. It will further assumed that the variance $\langle \eta^2\rangle = \lambda$, where clearly $\lambda >0$.

\begin{remark} 
Clearly, the choice of the value function  \fer{vf} is only one of the possible choices. For example, to differentiate the percentage of increasing the service time from the percentage of decreasing it, and to outline the difficulty to shorten it, we can consider the value function
 \be\label{diff}
  \Psi(s) = \gamma \frac{s^\delta -1}{\nu s^\delta +1} , \quad  s \ge 0,
 \ee
where the constant $ \nu >1$. In this case, \fer{bounds} modifies to
 \be\label{bound2}
 -\gamma \le \Psi(s) \le  \frac \gamma{\nu} < \gamma.
  \ee
In this case, the possibility to decrease the service time when it is too high is slowed down. As we shall see, this choice will modify the steady state distribution.
\end{remark}
Given the time service microscopic variation \fer{coll}, the study of the
time-evolution of the distribution of the service times density produced by 
interactions  of type \fer{coll}  can be obtained by
resorting to kinetic collision-like models \cite{Cer,PT13}, where the variation of the  density $f(w,t)$  obeys to a
Boltzmann-like equation. This equation is usually written
in weak form. It corresponds to say that the solution $f(w,t)$
satisfies, for all smooth functions $\varphi(w)$ (the observable quantities)
 \begin{equation}
  \label{kin-w}
 \frac{d}{dt}\int_{\R_+}\varphi(w)\,f(w,t)\,dx  = \frac 1\tau
  \Big \langle \int_{\R_+} \bigl( \varphi(w_*)-\varphi(w) \bigr) f(w,t)
\,dw \Big \rangle.
 \end{equation}
 Here expectation $\langle \cdot \rangle$ takes into account the presence of the random parameter $\eta$ in \fer{coll}. The positive constant $\tau$ measures the interaction frequency.

Clearly, the right-hand side of equation \fer{kin-w} represents a balance between the amount of service times for which agents change their size from $w$ to $w_*$ (loss term with negative sign) and the amount of service times for which agents change their service time from any other size $w_*$ to the actual service time  $w$  (gain term with positive sign).

%%%%%%%%%%%%%%%%%%%%%%%%%%%%%%%%%%%%%%%%%%%%%%%%%%%%%%%%%
\section{Quasi-invariant service time limit and Fokker-Planck equation}\label{quasi}

%%%%%%%%%%%%%%%%%%%%%%%%%%%%%%%%%%%%%%%%%%%%%%%%%%%%%%%%%%%%%%%

In reason of the nonlinearity (in the service time variable $w$) of the interaction \fer{coll}, the only conserved quantity of equation \fer{kin-w} is obtained by setting $\varphi = 1$. This conservation law implies that the solution to \fer{kin-w} remains a probability density for all subsequent times $t >0$.  The evolution of other moments is difficult to follow. As main example, we take $\varphi(w) = w$, which allows to obtain that the evolution of the mean value
 \[
 m(t) = \int_{\R_+}w\, f(w, t)\, dw.
  \]
Since
 \[
 \langle w_* - w \rangle = \gamma \frac{w^\delta -\bar w_L^\delta}{w^\delta +\bar w_L^\delta}\, w,
  \]
we obtain
 \be\label{evo-m}
 \frac{d }{dt}m(t) = \frac\gamma\tau\int_{\R_+} \frac{w^\delta -\bar w_L^\delta}{w^\delta +\bar w_L^\delta}\, w\, f(w,t)\, dw.
 \ee
Note that equation \fer{evo-m} is not explicitly solvable.  However, in view of condition \fer{bounds} the mean value of the solution to equation \fer{kin-w} remains bounded at any time $t >0$, provided that it is bounded initially, with the explicit upper bound
 \[
 m(t) \le m_0\exp \left\{\frac\gamma\tau \, t \right\}.
  \]
Analogous result holds for the second moment, which corresponds to assume $\varphi(w) = w^2$. In this case, since
 \[
 \langle w_*^2 -w^2\rangle = \big[ \Phi\left( w/\bar w_L \right)^2 -2 \Phi\left( w/\bar w_L\right) + \lambda \big] w^2 \le [\gamma^2 + \lambda]w^2
  \]
the boundedness of the initial second moment implies the boundedness of the second moment of the solution at any subsequent time $t>0$, with the explicit upper bound
\[
 m_2(t) \le m_{2,0}\exp \left\{\frac{\gamma^2 +\lambda}\tau \, t \right\}.
  \]
Let us suppose now that the interaction \fer{coll} is such that the service time is modified only of a very small amount. This can be easily achieved by setting for some value $\e$, with $\e \ll 1$ 
 \be\label{scal}
\Phi(s) \to \Phi_\e(s) = \gamma \frac{s^\e -1}{s^\e +1}, \quad \eta \to \sqrt\e \eta.
 \ee
Then, since for any time $t >0$ we can write \fer{evo-m} as
 \[
\frac{d }{dt}m(t) = \e \bar w_L \, \, \frac\gamma\tau\int_{\R_+} \frac 1\e\left[ \left(\frac w{\bar w_L}\right)^\e -1\right] \, \frac w{\bar w}\, \frac 1{\left(w/\bar w_L\right)^\e + 1}\, f(w,t)\, dw, 
 \]
 and, for $s \ge 1$, independently of the value of the small parameter $\e$
 \[
 \frac 1\e\left[ s^\e -1\right]  \le s,
  \]
 while for $s \le 1$
 \[
  \frac 1\e\left[ s^\e -1\right]  s \ge -1,
 \] 
for any given fixed time $t >0$ we can choose $\e$ in such a way to have a small variation of the mean value $m(t)$.  In this situation it is clear that, if we want to observe an evolution of the average value independent of $\e$,  we can resort to a scaling of the frequency $\tau$. If we set $\tau \to \e \tau $, and $f_\e(w, t) $ will denote the corresponding density, then  the evolution of the average value for $f_\e(w, t)$ satisfies
\[
\frac{d}{d\t}\int_{\R_+}w \,f_\e(w,t)\,dw  =  \bar w_L\, \, \frac\gamma\tau\int_{\R_+} \frac 1\e\left[ \left(\frac w{\bar w_L}\right)^\e -1\right] \, \frac w{\bar w_L}\, \frac 1{\left(w/\bar w_L\right)^\e + 1}\, f_\e(w,t)\, dw, 
 \]
namely a computable evolution law for the average value of $f$, which remains bounded even in the limit $\e \to 0$, 
since pointwise
 \be\label{AA}
 A_\e (w) = \frac 1\e\left[ \left(\frac w{\bar w_L}\right)^\e -1\right] \, \frac 1{\left(w/\bar w_L\right)^\e + 1} \to \frac 12 \log \frac w{\bar w_L}.
  \ee
 The reason is clear. If we assume that the interactions are scaled to produce a very small change in the service time, to observe  evolution of the average value independent of the smallness, we need to suitably increase the frequency of services.
By using the same scaling \fer{scal} one can easily obtain the evolution equation for  the second moment of $f_\e(w,t)$, which will be well-defined also in the limit $\e \to 0$ (cf. the analysis in \cite{FPTT}).  
%%%%%%%%%%%%%%%%%%%%%%%%%%%%%%%%%%%%%%%%%%

The previous discussion about moments enlightens  the main motivations and the mathematical ingredients that justify the passage from the kinetic model \fer{kin-w} to its continuous counterpart given by a Fokker--Planck equation. 
Given a smooth function $\varphi(w)$, let us expand in Taylor series $\varphi(w_*)$ around $\varphi(w)$. Using the scaling \fer{scal} one obtains
 \[
 %\label{cor2}
\langle w_* -w \rangle = - \e \, \gamma \,A_\e(w)\,w;  \quad  \langle (w_* -w)^2\rangle =  \left(\e^2 \, \gamma^2\, A_\e^2(w) + \e \lambda\right) w^2.
 \]
Therefore, in terms of powers of $\e$,  we easily obtain the expression
 \[
 %\label{tay}
\langle \varphi(w_*) -\varphi(w) \rangle =  \e \left( - \varphi'(w)\frac \gamma 2\,w \log \frac w{\bar w_L}
  + \frac \lambda 2 \, \varphi''(w)  w^2 \right) + R_\e (w),
 \]
where the remainder term $R_\e$, for a suitable $0\le \theta \le 1$,  given by 
 \begin{equations}\label{rem}
R_\e(w) = & \frac 12 \e^2 \, \varphi''(w) A_\e^2(w)\, w^2  + \e \gamma \left( A_\e(w) - \frac 12 \log \frac w{\bar w_L}\right)\,w +
\\  
& \frac 16  \langle \varphi'''(w+\theta(w_* -w))  (w_* -w)^3\rangle, 
 \end{equations}
 is such that 
  \[
  \frac 1\e \, R_\e(w) \to 0
  \]
 as  $\e \to 0$. Therefore, if  we set  $\tau \to \e \tau$,  we obtain that the evolution of the (smooth) observable quantity $\varphi(w)$ is given by
\[
\begin{aligned}
%  \label{m-l23}
 & \frac{d}{dt}\int_{\R_+}\varphi(w) \,f_\e(w,t)\,dw  = \\
 & \int_{\R_+} \left( - \varphi'(w)\,  \frac \gamma 2 \, w\, \log \frac w{\bar w_L} + \frac \lambda 2 \varphi''(w) w^2 \right) f_\e(w,t)\, dw \ + \frac 1\e \mathcal R_\e(w,t) ,
 \end{aligned}
 \]
where 
\[
\label{rem3}
\mathcal R_\e(t) = \int_{\R_+ } R_\e(w)  f_\e(w,t)\, dw,
\]
and $R_\e$ is given by \fer{rem}. Letting $\e \to 0$ shows that in consequence of the scaling \fer{scal} the weak form of the kinetic model \fer{kin-w} is well approximated by the weak form of a linear Fokker--Planck equation (with variable coefficients)
\begin{equations}
  \label{m-13}
 & \frac{d}{dt}\int_{\R_+}\varphi(w) \,g(w,t)\,dw  = \\
  & \int_{\R_+} \int_{\R_+} \left( - \varphi'(w) \frac \gamma 2 \, w\, \log \frac w{\bar w_L} + \frac \lambda 2 \varphi''(w) w^2 \right) g(w,t)\, dw 
   \end{equations}
Provided the boundary terms produced by the integration by parts vanish,  equation \fer{m-13} coincides with the weak form of the Fokker--Planck equation
 \begin{equation}\label{FP2}
 \frac{\partial g(w,t)}{\partial t} = \frac \lambda 2 \frac{\partial^2 }{\partial w^2}
 \left(w^2 g(w,t)\right )+ \frac \gamma 2
 \frac{\partial}{\partial w}\left(  w\, \log \frac w{\bar w_L} g(w,t)\right).
 \end{equation}
Equation \fer{FP2} describes the evolution of the distribution density $g(w,t)$ of service times $w \in \R_+$, in the limit of the quasi-invariant service time variations.  As often happens with Fokker-Planck type equations, the steady state density can be explicitly evaluated, and it results to be a lognormal density, with parameters linked to the details of the microscopic time service variation \fer{coll}.
 
 %%%%%%%%%%%%%%%%%%%%%%%%%%%%%%%%%%%%%%%%%%%%%%%%%%%%%%%%%%%%%%%%%%%
\section{The steady state is a lognormal density}

%%%%%%%%%%%%%%%%%%%%%%%%%%%%%%%%%%%%%%%%%%%%%%%%%%%%%%%%%%%%%%%%%%%%%%%%%
The stationary distribution of the Fokker--Planck equation \fer{FP2} is easily found by solving the differential equation 
 \be\label{sd}
 \frac \lambda 2 \frac{d }{dw}
 \left(w^2 g(w)\right )+ 
 \frac \gamma 2  w\, \log \frac w{\bar w_L}\, g(w)   =0.
 \ee
Solving \fer{sd} with respect to $h(w)= w^2 g(w)$ by separation of variables gives as unique solution to \fer{sd} of unit mass the density
 \be\label{equilibrio}
g_\infty(w) = \frac 1{\sqrt{2\pi \sigma}\, w} 
\exp\left\{ - \frac{(\log w - \mu)^2}{2 \sigma}\right\},
 \ee 
where 
 \be\label{para}
 \sigma = \frac \lambda\gamma,  \quad \mu = \log \bar w_L - \sigma.
 \ee
 Hence, the equilibrium distribution \fer{equilibrio} takes the form of a lognormal density with mean and variance given respectively by
 \be\label{moments}
 m(g_\infty) = \bar w_L e^{-\sigma/2}, \quad Var(g_\infty) = \bar w_L^2 \left( 1 - e^{-\sigma}\right).
 \ee
Note that the moments are expressed in terms of the parameters $\bar w_L$, $\lambda$ and $\gamma$ denoting respectively the  service time limit $\bar w_L$, the variance $\lambda$ of the random effects and the size $\gamma$ of the value function $\phi$. 

In particular, both the mean value and the variance of the steady state density depend only on the ratio $\sigma = \lambda/\gamma$  between the variance $\lambda$ of the random percentage of service time allowed, and the maximal percentage allowed of possible variation of service time.

If the value $\sigma$ satisfies 
 \be\label{www}
 \sigma \ge 2 \log \frac{\bar w_L}{\bar w},
 \ee 
the mean of the service time is lower that the fixed ideal time $\bar w$, which represents a very favorable situation for the call center. 

\begin{remark} 
If the value function \fer{diff} is considered, then \fer{AA} will be substituted by
 \be\label{AB}
 A_\e (w) = \frac 1\e\left[ \left(\frac w{\bar w_L}\right)^\e -1\right] \, \frac 1{\nu \left(w/\bar w_L\right)^\e + 1} \to \frac 1{1+\nu}  \log \frac w{\bar w_L}.
  \ee
 where the constant $ \nu >1$. In this case, the drift term in the Fokker--Planck equation \fer{FP2} modifies to
 \be\label{drift2}
D(g)(w) =   \frac \gamma {1+\nu}
 \frac{\partial}{\partial w}\left(  w\, \log \frac w{\bar w_L} g(w,t)\right)
   \ee
In this case, by setting
 \[
 \tilde \gamma =  \gamma \, \frac 2 {1+\nu} < \gamma,
  \]
the steady state \fer{equilibrio} remains a lognormal density, with $\sigma$ substituted by $\tilde\sigma = \sigma(1+\nu)/2 > \sigma$. 
\end{remark}

%%%%%%%%%%%%%%%%%%%%%%%%%%%%%%%%%%%%%%%%%%%%%%%%%%%%%%%%%%%%%%%%%%%%%%%%%%%%
\section{Numerical experiments}
\label{numerica}

The mathematical analysis of the Fokker--Planck equation \fer{FP2} will be discussed elsewhere. The main goal of this paper was the statistical approach to the duration of time services in terms of microscopic interactions in a  multi-agent system. Consequently, it is of primary importance in this modeling process to verify the coherence of the evolution of the density towards a steady profile with certain characteristics with the set of allowed data. 
 
In this section, we report our numerical experiments using the data provided by our industrial partner during January 2018, relative to a call center that manages back office practices, here simply denoted as jobs. In such call centers,  more than 300 operators work every day a number of jobs
that ranges between 10'000 and 20'000. The jobs are classified into 270 different job types. 
Every job type has different Quality of Service (QoS) constraints, among which, the most important is the maximum service time duration $\bar{w}_L$.
In addition, as described in Section 2, each operator gets from the service manager an ideal service time $\bar{w}$ for every job type.

\begin{figure}[tp!]
\centering
\includegraphics[width=0.9\textwidth]{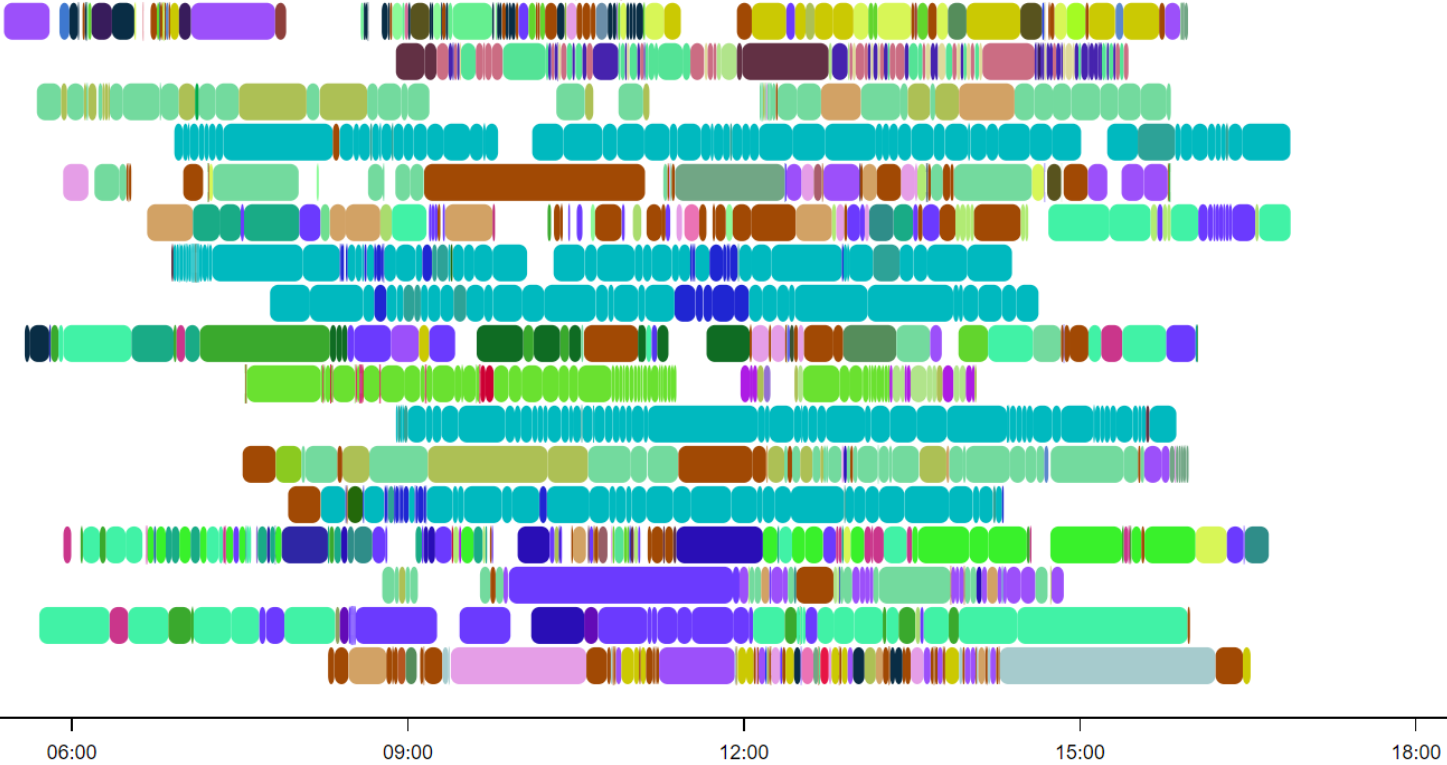}
\caption{Example of a working day in call center for 17 operators: each row represents the sequence of jobs served by an operator.
Each small rectangle represents the elaboration of a single job.
The colors represent the different job types.} \label{fig:1}
\end{figure}

Figure \ref{fig:1} shows a snapshot of a working day of 17 operators of the call center, where the horizontal axis represents the day time, from 6:00 am to 6:00 pm. 
The plot has a row (sequence) of rectangles for each operator, and the width of each rectangle is proportional to the service time duration.
The different colors are used to highlight the job type (rectangles with the same colors refer  to the same job type).
For instance, the first operator (first row) worked several different job types, while the fourth operator (fourth row) worked almost always the same type of job.

\begin{figure}[tph!]
\centerline{\includegraphics[width=0.8\textwidth]{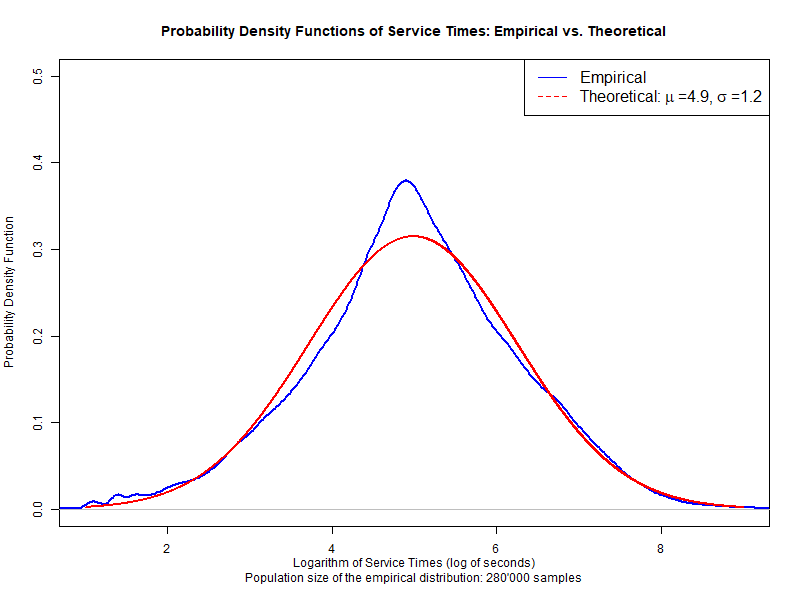}}
    \caption{}
    \label{fig2}
\end{figure}

Regardless of the job type, we are first interested in studying the distribution of the service time (measured in seconds) of every job, that is, 
in a sense, the width of every single rectangle. We consider the service time as the observations of the distribution we want to study.
By looking at Figure \ref{fig:1}, it is hard to understand whether the longer jobs (larger rectangles)
should be considered as simple outliers, or if they should be regularly included in the distribution.
However, if we change perspective, and instead of looking at the service time, we just consider the logarithm of the service time, we get a clearer picture. 
For instance, by plotting the empirical probability density function of the log of the service time (i.e., $\log(w)$) for the
whole data set, which consists of more than 280'000 samples of service time, we obtain Figure \ref{fig2}.
Indeed, Figure \ref{fig2} compares the empirical distribution (blue line) obtained with the empirical data, 
with the fitted theoretical lognormal distribution (red line), which has mean $\mu=4.9$ and variance $\sigma=1.2$.
Note that in the horizontal axis we have the logarithm of the service time, and, hence, to the value 8 corresponds a time of $e^8\approx 3000$ seconds. 

\begin{figure}[tph!]
\centerline{\includegraphics[width=\textwidth]{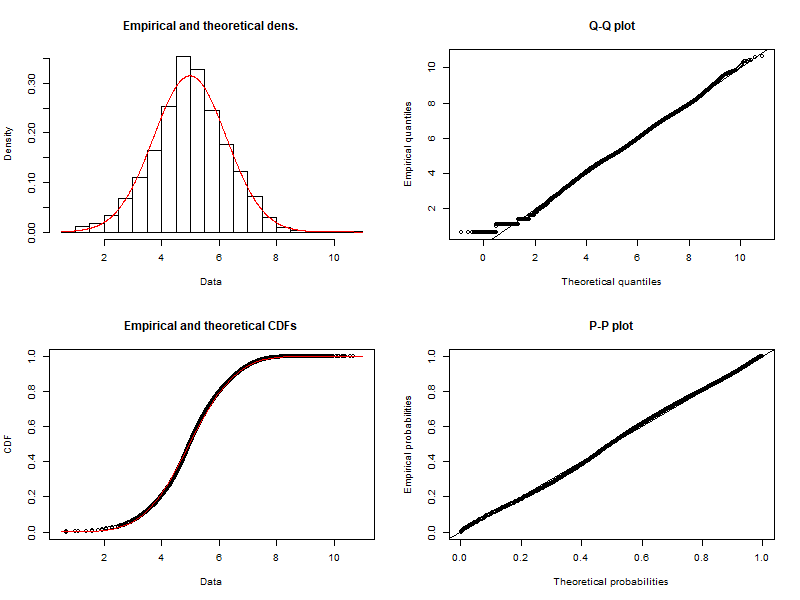}}
    \caption{Evaluation of fitting the log of the service time $log(w)$ with a Gaussian distribution.}
    \label{fig3}
\end{figure}

The parameters of the lognormal distribution plotted in red in Figure \ref{fig2} are estimated by maximizing the likelihood function,
using the {\it fitdist} package of the R statistical software. As a first step, we compute the log of the service time $\log(w)$ for the whole
set of 280'000 samples provided by our industrial partner.
Figure \ref{fig3} shows the quality of fitting $\log(w)$ with a Gaussian distribution. The four subplots show in order:
(i) the histogram of the empirical observations along with the kernel density (red line); 
(ii) the quantile-quantile plot (Q-Q plot) which compares the empirical quantiles with the theoretical quantiles;
(iii) the empirical and the theoretical empirical Cumulative Distribution Functions (CDF);
(iv) the probability-probability plot (P-P plot) which compares the probabilities quantiles with the theoretical probabilities.
Remark that in particular the Q-Q plot and the P-P plot clearly show the goodness of our fitting, since in both cases the data follow a straight line.

Once verified that the service time distributions follow a lognormal distribution, we have performed a deeper analysis by fitting
a lognormal distribution for each job type. This analysis has the objective to observe how the distribution of the service times behaves
with respect of the ideal time $\bar{w}$, given by the service manager, and the limit $\bar{w}_L$, given by the QoS time constraints, which are different for each job type.
As in the previous figures, the blue empirical density function refers to the real data, and the red probability density function shows
the lognormal distribution with the mean $\mu$ and deviations $\sigma$ detailed in each subfigure. 
In addition, the plots show with vertical lines the values of $\log{(\bar{w})}$ and $\log{(\bar{w}_L)}$.
For instance, the first subplot, which refers to Job Type 1, shows the distribution of 28'425 service times. The blue dotted vertical line refers
to the log of the ideal time $\log{(\bar{w})=6}$ (i.e., 400 seconds), and 
the dashed green vertical line refer to the log of the time limit $\log{(\bar{w}_L)=7.3}$ (i.e., 1'500 seconds).
The fitted lognormal distribution has mean $\mu=4.9$ and variance $\sigma=1.2$.

\begin{figure}[tph!]
\centerline{\includegraphics[height=\textheight]{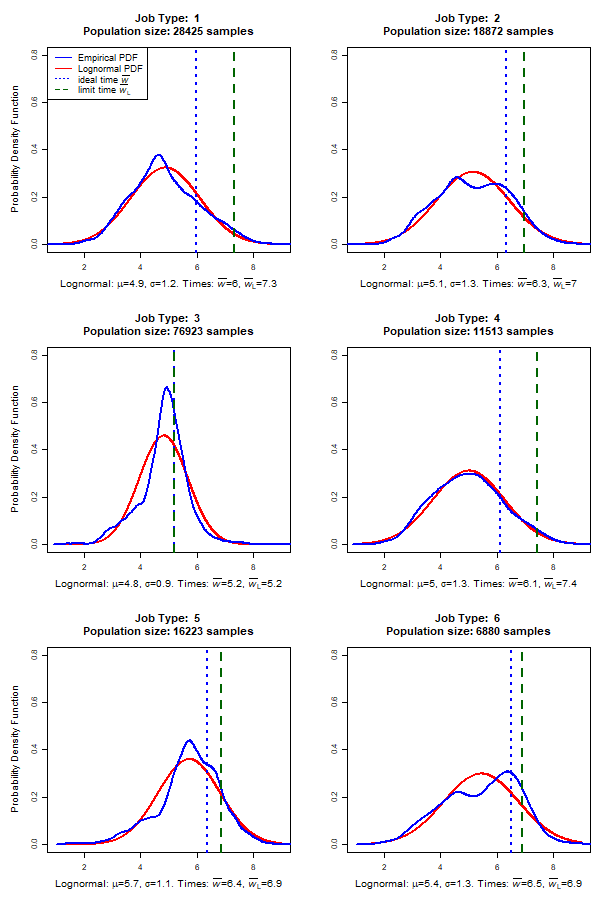}}
    \caption{Fitted lognormal distributions over the six most frequent job types. The vertical dashed lines illustrate the reference times $\bar{w}_L$ for the Quality of Service constraint.}
    \label{fig4}
\end{figure}

From the plots of Figure \ref{fig4}, it is clear that while both reference times plays a role in the shape of the final distributions of service times,
the time limit $\bar{w}_L$ puts a stronger pressure on the operator, since, except for Job Type 3, only a small percentage of service durations exceed such limit.
Note that Job Type 3 is different from the others, since its ideal time is equal to the time limit, which is $e^{5.2}\approx 180$ seconds.
It is indeed hard to image a successfully completion of a service call in less than three minutes.

%%%%%%%%%%%%%%%%%%%%%%%%%%%%%%%%%%%%%%%%%%%%%%%%%%%%%%%%%%%%%%%%%%%%%%%%%%%%%%%%%
\section{Conclusions}
In this note we tried to explain the numerical evidence of a lognormal distribution in the service time duration in a call center, by resorting to classical methods of kinetic theory and statistical physics. We described the process of time duration in terms of microscopic interactions in a multi-agent system in which agents try to adapt their service time by resorting to a suitable value function, in the spirit of prospect theory of Kahneman and Twersky \cite{KT}.  Our analysis allows to describe the whole process in terms of the evolution of a probability density satisfying a Fokker--Planck type equation with a lognormal distribution as stationary state. In analogy with the classical kinetic theory of rarefied gases \cite{Cer}, convergence towards the stationary solution takes a very short time. Consequently, what we see from the numerical analysis of data is essentially the lognormal equilibrium density.

%%%%%%%%%%%%%%%%%%%%%%%%%%%%%%%%%%%%%%%%%%%%%%%%%%%%%%%%%%%%%%%%%%%%%%%%%%%%%%%

\section*{Acknowledgement} This work has been written within the
activities of GNFM group  of INdAM (National Institute of
High Mathematics), and partially supported by  MIUR project ``Optimal mass
transportation, geometrical and functional inequalities with applications''.
The data have been made available within the industrial project entitled {\it Resource Optimization Allocation},
funded by ComData via the Mathesia crowd sourcing web platform\footnote{\url{http://www.mathesia.com}}. 
The authors thanks Adriana Mina, Andrea Bonomo and Daniele Medone, 
from ComData and DeltaProgetti2000 for the kind support in explaining the way of operating of a modern call center, and for providing the raw data used for the numerical experiments.

%%%%%%%%%%%%%%%%%%%%%%%%%%%%%%%%%%%%%%%%%%%%%%%%%%%%%%%%%%%%%%%%%%%%%%%%%%%%%%%

\end{document}